\documentclass[aps,prl,floatfix,twocolumn,footinbib,amsmath,amssymb]{revtex4}
\usepackage{amsmath,amsthm,amssymb,color,psfrag}
\usepackage{url}
\usepackage{latexsym}
\usepackage{graphicx}
\usepackage{slashed}

\definecolor{darkred}{rgb}{0.6,0.0,0.0}
\definecolor{darkblue}{rgb}{0.0,0.0,0.5}
\definecolor{darkgreen}{rgb}{0.0,0.5,0.0}
\definecolor{brown}{rgb}{0.0,0.0,0.0}

\newcommand{\be}{\begin{equation}}
\newcommand{\ee}{\end{equation}}
\newcommand{\bea}{\begin{eqnarray}}
\newcommand{\eea}{\end{eqnarray}}

\begin{document}

\topmargin -0.5in

\title{Probing the hardest branching of jets in heavy ion collisions}
\author{Yang-Ting Chien$^{a,b}$ and Ivan Vitev$^a$}

\affiliation{
$^a$ Theoretical Division, T-2, Los Alamos National Laboratory, Los Alamos, NM 87545, USA\\
$^b$ Erwin Schr{\" o}dinger International Institute for Mathematical Physics, Universit{\"a}t Wien, Wien, Austria
}

\begin{abstract}
We present the first calculation of the momentum sharing and angular separation distributions between the leading subjets inside a reconstructed jet in heavy ion collisions. These observables are directly sensitive to the hardest branching in the process of jet formation and are, therefore, ideal for studying the early stage of the in-medium parton shower evolution. The modification of the momentum sharing and angular separation distributions in lead-lead relative to proton-proton collisions is evaluated using the leading-order medium-induced splitting functions obtained in the framework of soft-collinear effective theory with Glauber gluon interactions.
Qualitative and in most cases quantitative agreement between theory and preliminary CMS measurements suggests that the parton shower in heavy ion collisions can be dramatically modified early in the branching history. We propose a new measurement which will illuminate the angular distribution of the hardest branching within jets in heavy ion collisions.
\end{abstract}
\maketitle

The dramatic suppression of hadron and jet cross sections observed at the Relativistic Heavy Ion Collider
(RHIC) \cite{Adcox:2001jp,Adler:2002xw,Adcox:2004mh,Arsene:2004fa,Back:2004je,Adams:2005dq} and the Large Hadron Collider (LHC)
\cite{Aad:2010bu,Aamodt:2010jd,Chatrchyan:2011sx,CMS:2012aa,Aad:2012vca,Aad:2014bxa,CMS:prelim,Adam:2015ewa}
signals the strong modification of parton showers within  strongly-interacting matter. This jet quenching phenomenon has been an essential tool to study the properties of the quark-gluon plasma (QGP) produced in ultrarelativistic nucleus-nucleus (A+A) collisions. The emergence of the in-medium parton branching, qualitatively different from the one which
governs the jet formation in $e^++e^-$, $e^++p$, $p+p$ collisions, is at the heart of all jet modification studies.
Although the traditional energy loss picture has been very successful in describing the suppression of cross section, to disentangle the detailed jet formation mechanisms in the medium requires comprehensive studies of jet substructure observables.

In the past few years there has been a proliferation of jet substructure measurements in A+A collisions~\cite{Chatrchyan:2013kwa,Chatrchyan:2014ava,Aad:2014wha,CMS:2016jys}, which gave differential and correlated information about how quark and gluon radiation is redistributed due to medium interactions. It is now definitively established that the jet shape~\cite{Ellis:1992qq} and the jet fragmentation function~\cite{Procura:2009vm}, which describe the transverse and longitudinal momentum distributions inside jets, are modified in
heavy ion collisions. Both of these observables depend strongly on the partonic origin of jets, and their nontrivial modification patterns are partly due to the increase of the quark jet fraction in heavy ion collisions~\cite{Chien:2014nsa,Chien:2015hda,Chien:2015ctp,medium_jet_frag}. To better understand the jet-by-jet modifications for these observables, one can devise strategies to isolate purer quark or gluon jet samples.

\begin{figure}
\centering
\includegraphics[width=0.45\textwidth]{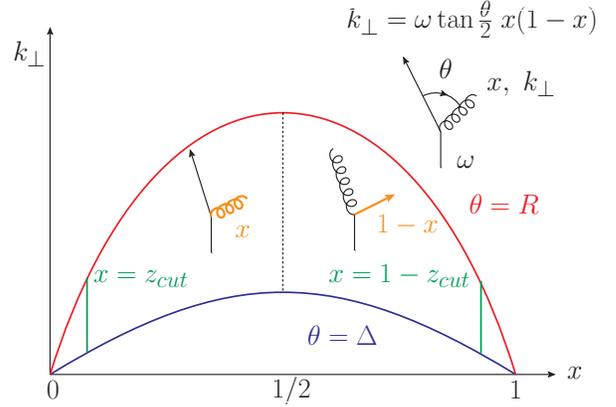}
\caption{ Illustration of the phase space regions for the $z_g$ distribution calculation constrained by $R$, $\Delta$ and $z_{cut}$. At leading order a collinear parton splits into partons with momenta $k=(x\omega,k_\perp^2/x\omega,k_\perp)$ and $p-k$. Depending on the kinematics of the splitting, the soft branch can be either of the partons.}
\label{PSintegral}
\end{figure}

Another collinear type of jet substructure observable, called the groomed momentum sharing, has been studied in the context of the soft drop jet grooming procedure~\cite{Larkoski:2014wba} and Sudakov safety~\cite{Larkoski:2015lea}. This observable probes the hard branching in the jet formation and is dominated by the leading-order Altarelli-Parisi splitting functions~\cite{Altarelli:1977zs}. Given a jet reconstructed using the anti-$k_T$ algorithm~\cite{Cacciari:2008gp} with radius $R$, one reclusters the jet using the Cambridge-Aachen algorithm~\cite{Dokshitzer:1997in,Wobisch:1998wt} and goes through the clustering history, grooming away the soft branch at each step until the following condition is satisfied,
\be
    z_{cut} < \frac{\min(p_{T_1},p_{T_2})}{p_{T_1}+p_{T_2}} \equiv z_g\;,
\ee
i.e., the soft branch is not carrying less than a $z_{cut}$ fraction of the sum of the transverse momenta therefore is not dropped. Note that by definition $z_{cut}<z_g<\frac{1}{2}$. Due to detector granularity one also demands that the angular separation between the two branches be greater than the angular resolution $\Delta$,
\be
    \Delta < \Delta R_{12} \equiv r_g\;.
\ee
More generally, by selecting the angular separation $\Delta R_{12}$, defined as the groomed jet radius $r_g$, one could also examine the momentum sharing distribution $p(z_g)$ at different splitting angles and the $p(r_g)$ distribution.

For jets with small radii~\cite{Dasgupta:2014yra,Chien:2015cka,Becher:2015hka,Kang:2016mcy}, the $z_g$ distribution can be described by the collinear parton splitting functions. At leading order, for a parton $i$ with collinear momentum $p=(\omega,0,0)$~\footnote{Note that we use lightcone coordinates, i.e. $\omega = 2 p_T$ for midrapidity jets with transverse momentum $p_T$.} splitting into partons $j,l$ with momenta $k=(x \omega,k_\perp^2/x \omega,k_\perp)$ and $p-k$, the splitting functions in vacuum ${\cal P}^{vac}_{i\rightarrow jl}(x,k_\perp)$ are well-known and their non-singular parts are reproduced below,
\bea
    {\cal P}^{vac}_{q\rightarrow qg} &=& \frac{\alpha_s(\mu)}{\pi}C_F\frac{1+(1-x)^2}{x}\frac{1}{k_\perp} \, , \\
    {\cal P}^{vac}_{g\rightarrow gg} &=& \frac{\alpha_s(\mu)}{\pi}C_A\Big[\frac{1-x}{x}+\frac{x}{1-x}+x(1-x)\Big]\frac{1}{k_\perp} \, , \quad \\
    {\cal P}^{vac}_{g\rightarrow q\bar q} &=& \frac{\alpha_s(\mu)}{\pi}T_Fn_f\Big[x^2+(1-x)^2\Big]\frac{1}{k_\perp}\, , \quad \\
 {\cal P}^{vac}_{q\rightarrow gq} &=&  {\cal P}^{vac}_{q\rightarrow qg}(x\rightarrow 1-x) \, .
\eea
The $z_g$ distribution is calculated by integrating the splitting functions over the partonic phase space constrained by $R$, $\Delta$ and $z_{cut}$  and shown in FIG. \ref{PSintegral},
\be
    p_i(z_g)=\frac{\int_{k_\Delta}^{k_R}dk_\perp\overline{\cal P}_i(z_g,k_\perp)}{\int_{z_{cut}}^{1/2}dx\int_{k_\Delta}^{k_R}dk_\perp\overline{\cal P}_i(x,k_\perp)}\; .
    \label{pzg}
\ee
Here, $k_\Delta = \omega x(1-x)\tan{\frac{\Delta}{2}}$, $k_R = \omega x(1-x)\tan{\frac{R}{2}}$ and
$$\overline{\cal P}_i(x,k_\perp) = \sum_{j,l}\Big[{\cal P}_{i\rightarrow j,l}(x,k_\perp)+{\cal P}_{i\rightarrow j,l}(1-x,k_\perp)\Big]. $$
Note that for anti-$k_T$ jets the angle $\theta$ between the two final state partons satisfies $\Delta < \theta < R$. The effect of running coupling can be taken into account by setting $\mu=k_\perp$ in the splitting function. The final $z_g$ distribution is then weighted by the jet production cross sections,
\be
    p(z_g)=\frac{1}{\sigma_{\rm total}}\sum_{i=q,g}\int_{PS} d\eta dp_T \frac{d\sigma^{i}}{d\eta dp_T} p_i(z_g)\;,
    \label{pzgX}
\ee
with the phase space cuts ($PS$) on the jet $p_T$ and $\eta$ imposed in experiments.

The $z_g$ distributions for quark-initiated  and gluon-initiated jets are very similar throughout the whole $z_g$ region. The color factors $C_F=4/3$ and $C_A=3$ for quarks and gluons cancel and the distributions follow approximately $1/z_g$, the leading behavior of the splitting functions in Eqs. (3) and (4) for $x < 1/2$.  The insensitivity of $z_g$ to the partonic origin of jets implies that its modification in heavy ion collision is not significantly affected by the change of the quark/gluon jet fraction as one observes in the jet shape and the jet fragmentation function.

In the presence of the medium,
\begin{equation}
    {\cal P}_{i\rightarrow jl}(x, k_\perp)={\cal P}^{vac}_{i\rightarrow jl}(x, k_\perp)+{\cal P}^{med}_{i\rightarrow jl}(x, k_\perp)\;,
\end{equation}
which is the sum of the vacuum and medium-induced splitting functions. The later were calculated using soft-collinear effective theory \cite{Bauer:2000ew, Bauer:2000yr, Bauer:2001ct, Bauer:2001yt, Bauer:2002nz} with Glauber gluon interactions ($\rm SCET_G$) \cite{Idilbi:2008vm, Ovanesyan:2011xy,Ovanesyan:2011kn,Fickinger:2013xwa} in a QGP model consisting of thermal quasi-particles undergoing longitudinal Bjorken-expansions \cite{Bjorken:1982tu}. $\rm SCET_G$ is an effective field theory of QCD suitable for describing jets in the medium. It goes beyond the traditional parton energy loss picture in the soft gluon limit, and it provides a systematic framework for resumming jet substructure observables and consistently including medium modifications. The medium-induced splitting functions used in this paper have been previously applied to describe and predict several hadron and jet observables in heavy ion collisions~\cite{Kang:2014xsa,Chien:2015vja,Chien:2015hda,medium_jet_frag}.

It can be seen analytically and confirmed numerically that in the region of interest $x < 1/2$,
the leading behavior of the in-medium splitting functions follows approximately $1/x^2$ ~\cite{Ovanesyan:2011kn}.
A testable hypothesis is that the momentum sharing distribution will show enhancement at the smallest values of $z_g$ and
suppression near $z_g=1/2$.

With the full collinear parton splitting functions in the medium, Eqs. (\ref{pzg}) and (\ref{pzgX}) are completely general and can be used to calculate the momentum sharing distribution in heavy ion collisions. The jet cross section was calculated by incorporating the jet energy loss due to out-of-cone radiation, and the small cold nuclear matter effects as in~\cite{Vitev:2007ve,Chien:2015hda}. However, since $z_g$ is insensitive to the flavor of jet-initiating partons, the effect from the change of quark/gluon jet fractions due to the different amounts of cross section suppression is minor.

For the cross section calculations, we use the CTEQ5M parton distribution functions~\cite{Tung:2006tb} and the leading-order ${\cal O}(\alpha_s^2)$ QCD partonic cross section results. We estimate the theoretical uncertainty by varying the coupling between the jet and the QCD medium $g=2.0\pm0.2$ as in~\cite{Chien:2015hda}. We use the two-loop running of the strong coupling constant with $\alpha_s(m_Z)=0.1172$.

The great utility of the momentum sharing distribution in heavy ion collisions lies on the fact that, one can select the jet transverse momentum and the angle between the two leading subjets to ensure large splitting virtuality and, consequently, a branching which happens shortly after the hard scattering inside the QGP. Indeed, the branching time
 \begin{equation}
\tau_{\rm br} [{\rm fm}] = \frac{0.197\; {\rm GeV~fm} }{z_g(1-z_g) \,  \omega [{\rm GeV}]  \, \tan^2( r_g / 2)}
\end{equation}
suggests that for typical jets with $\omega = 2 p_T  = 400$~GeV, $r_g = 0.1$ and $z_g = 0.1$, the branching time $\tau_{\rm br}  < 2$~fm. This is  much smaller than the size of the QGP created in
Pb+Pb collisions at the LHC and allows us to test whether the medium modification of parton branchings happens early in the
shower evolution.

We compare our calculations to the preliminary data taken by the CMS collaboration at the LHC Run II at $\sqrt {s_{\rm NN}} = 5.02$ TeV~\cite{CMS:2016jys}. In both proton-proton and lead-lead (Pb+Pb) collisions, the jets are reconstructed using the anti-$k_T$ algorithm with $R=0.4$~\cite{Cacciari:2008gp}. They  are then groomed using the soft-drop jet grooming procedure~\cite{Larkoski:2014wba}. The parameters chosen in the CMS measurements are $\beta = 0$ and $z_{cut} = 0.1$. Another cut on $\Delta R_{12} > 0.1$ is imposed due to the detector resolution where $\Delta R_{12}$ is the distance between the two branches in the pseudorapidity-azimuthal angle plane. The requirement also effectively selects jets with the branching angle greater than $0.1$. The groomed momentum sharing $z_g$ and its normalized distribution
\be
    p(z_g)=\frac{1}{N_{\rm jet}}\frac{dN}{dz_g}\;,
\ee
are measured. The jets are selected with the following cuts on the jet transverse momentum ($p_T$) and pseudorapidity ($\eta$): $p_T > 140~{\rm GeV}$ and $|\eta|<1.3$. The in-medium momentum sharing modification is quantified by taking the ratio of the $z_g$ distributions in proton-proton and lead-lead collisions,
\begin{equation}
    R^{p(z_g)}_{AA}=p(z_g)^{PbPb}\big/ p(z_g)^{pp}\;.
\end{equation}
The modification patterns are examined across a wide range of $p_T$ bins with different collisional centralities.

\begin{figure}
\begin{center}
    \psfrag{x}{$z_g$}
    \psfrag{y}{$\frac{p(z_g)^{PbPb}}{p(z_g)^{pp}}$}
    \includegraphics[width=0.45\textwidth]{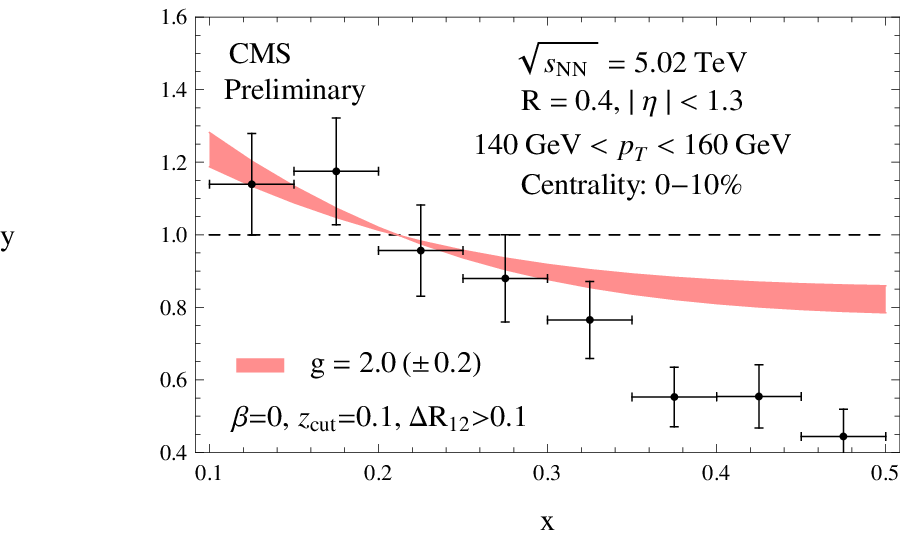}\\
    \includegraphics[width=0.45\textwidth]{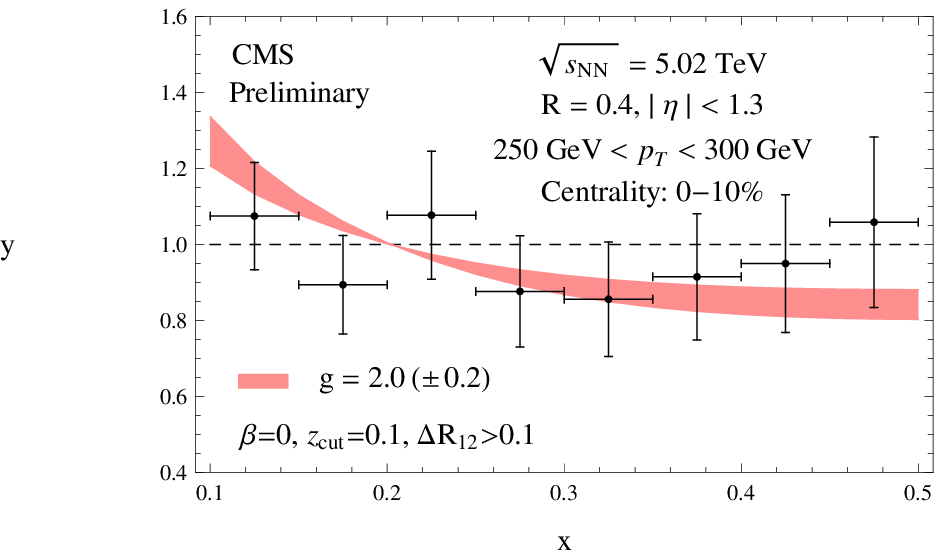}
\caption{Comparison of theoretical calculations and preliminary CMS data for the ratio of momentum sharing distributions of inclusive anti-$k_T$ $R=0.4$ jets in  central Pb+Pb and p+p collisions at $\sqrt {s_{\rm NN}} = 5.02$ TeV. Jets are soft-dropped with $\beta = 0$, $z_{cut} = 0.1$ and $\Delta R_{12} > 0.1$. Bands correspond to the theoretical uncertainty estimated by varying the coupling between the jet and the medium ($g=2.0\pm0.2$). Upper panel: modification for jets with $140~{\rm GeV}< p_T < 160~{\rm GeV}$ and $|\eta|<1.3$. Lower panel: modification for jets with $250~{\rm GeV}< p_T < 300~{\rm GeV}$ and $|\eta|<1.3$. }
\label{SplitpT}
\end{center}
\end{figure}

FIG. \ref{SplitpT} shows the result for the ratio of the momentum sharing distributions of inclusive jets in 0-10\% central Pb+Pb and p+p  collisions   at $\sqrt {s_{\rm NN}} = 5.02$ TeV. We consider two $p_T$ bins $140~{\rm GeV}< p_T < 160~{\rm GeV}$ (upper panel) and $250~{\rm GeV}< p_T < 300~{\rm GeV}$ (lower panel) to study the modification pattern as a function of the jet transverse momentum. The preliminary CMS data shows a strong modification of the momentum sharing distribution for jets with lower $p_T$ in central collisions, and the modification decreases quite quickly when the jet $p_T$ becomes higher. The red bands correspond to the theoretical calculations with the variation of $g=2.0\pm0.2$. We find that the modification does decrease as the jet $p_T$ increases. However, the $p_T$ dependence in our theory calculation is not as strong as suggested in the preliminary CMS measurements, with the amount of modification around $z_g=0.5$ underestimated in our calculation for lower $p_T$ jets. For jets with higher $p_T$, our calculation is consistent with the preliminary CMS data within the experimental uncertainties.

\begin{figure}
\begin{center}
    \psfrag{x}{$z_g$}
    \psfrag{y}{$\frac{p(z_g)^{PbPb}}{p(z_g)^{pp}}$}
    \includegraphics[width=0.45\textwidth]{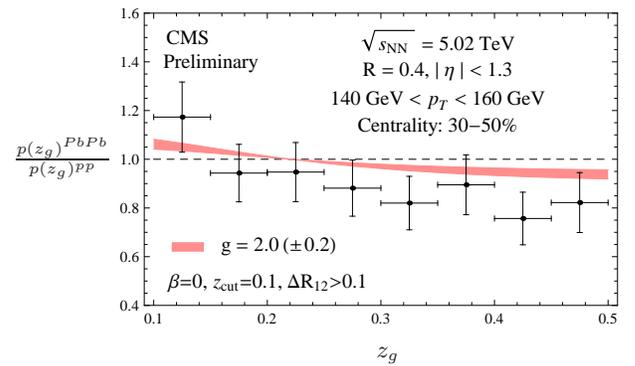}
\caption{Comparison of theoretical calculations and preliminary CMS data for the momentum sharing modification of inclusive jets in proton-proton and lead-lead collisions at $\sqrt {s_{\rm NN}} = 5.02$ TeV. Shown are the same studies as in FIG. \ref{SplitpT} for anti-$k_T$ $R=0.4$ jets with $140~{\rm GeV}< p_T < 160~{\rm GeV}$ and $|\eta|<1.3$ in mid-peripheral collisions. The same soft-drop parameters are used to groom the jets.}
\label{SplitMidperi}
\end{center}
\end{figure}

FIG. \ref{SplitMidperi} shows the modification of the momentum sharing distribution for inclusive jets in mid-peripheral lead-lead collisions with centrality 30-50\% at $\sqrt {s_{\rm NN}} = 5.02$ TeV. Here we only examine jets in the $140~{\rm GeV}< p_T < 160~{\rm GeV}$ bin since the modification is larger for lower $p_T$ jets. Both the CMS preliminary data and our calculation show moderate modifications of the $z_g$ distributions, and we are consistent with each other. The medium modification of the $z_g$ distribution decreases with collisional centrality.

\begin{figure}
\begin{center}
    \psfrag{x}{$z_g$}
    \psfrag{y}{$\frac{p(z_g)^{PbPb}}{p(z_g)^{pp}}$}
    \includegraphics[width=0.45\textwidth]{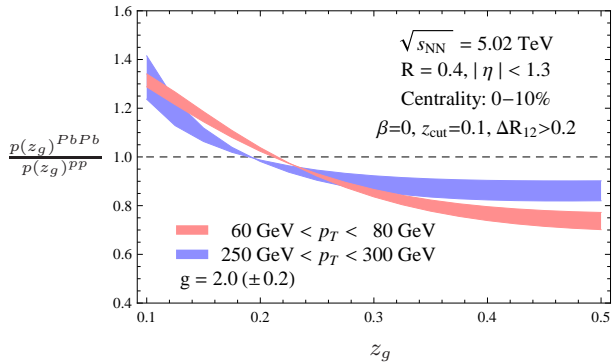}
\caption{Theoretical calculations for the momentum sharing distribution ratio of inclusive jets in proton-proton and central lead-lead collisions at $\sqrt {s_{\rm NN}} = 5.02$ TeV. Jets are soft-dropped with $\beta = 0$, $z_{cut} = 0.1$ and $\Delta R_{12} > 0.2$. We study its jet $p_T$ dependence and provide results for $60~{\rm GeV}< p_T < 80~{\rm GeV}$ (red band) and $250~{\rm GeV}< p_T < 300~{\rm GeV}$ (blue band).}
\label{SplitpTR12}
\end{center}
\end{figure}

Predictions for the momentum sharing distribution ratios for inclusive jets in proton-proton and central lead-lead collisions at $\sqrt {s_{\rm NN}} = 5.02$ TeV
are shown in FIG. \ref{SplitpTR12}.
We consider the $p_T$ bins $60~{\rm GeV}< p_T < 80~{\rm GeV}$ (red band) and $250~{\rm GeV}< p_T < 300~{\rm GeV}$ (blue band). However, whereas in the CMS preliminary measurements the cut $\Delta R_{12} > 0.1$ is imposed in the jet selection, here we present the theoretical calculations with a more stringent cut $\Delta R_{12} > 0.2$ to study the $z_g$ distribution with wider splitting angles. We find that the modification increases (decreases) with $\Delta R_{12}$ for low (high) $p_T$ jets, rendering a stronger $p_T$ dependence in the modification pattern. This can be understood because the angular scale set in the medium-induced splitting function should be proportional to the ratio between the medium temperature and the jet transverse momentum. By probing radiation with wider angles we should see that the medium modification decreases faster with the jet $p_T$.

\begin{figure}
\begin{center}
    \psfrag{x}{$r_g$}
    \psfrag{y}{$\frac{p(r_g)^{PbPb}}{p(r_g)^{pp}}$}
    \includegraphics[width=0.45\textwidth]{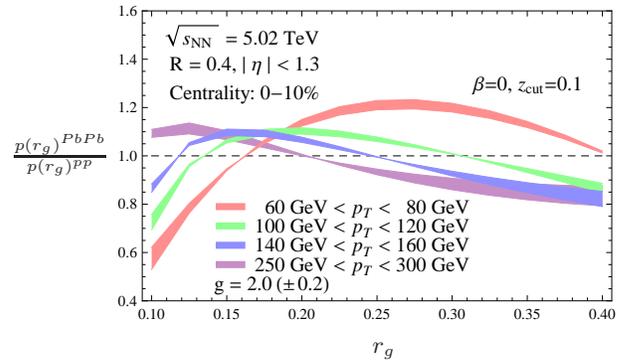}
\caption{Theoretical calculations for the groomed jet radius modification of inclusive jets in proton-proton and central lead-lead collisions at $\sqrt {s_{\rm NN}} = 5.02$ TeV. The soft drop parameters $\beta = 0$, $z_{cut} = 0.1$ and $\Delta R_{12} > 0.1$ are used. Shown are results for four $p_T$ bins with $60~{\rm GeV}< p_T < 80~{\rm GeV}$ (red band), $100~{\rm GeV}< p_T < 120~{\rm GeV}$ (green band), $140~{\rm GeV}< p_T < 160~{\rm GeV}$ (blue band) and $250~{\rm GeV}< p_T < 300~{\rm GeV}$ (purple band).}
\label{Radius}
\end{center}
\end{figure}

An important new observable that we propose to study in heavy ion collisions is the angular separation distribution $r_g \equiv \Delta R_{12}$ of the leading subjets inside a groomed jet. At leading order,
\be
    p_i(r_g)=\frac{\int_{z_{cut}}^{1/2}dx~p_Tx(1-x)\overline{\cal P}_i(x,k_\perp(r_g,x))}{\int_{z_{cut}}^{1/2}dx\int_{k_\Delta}^{k_R}dk_\perp\overline{\cal P}_i(x,k_\perp)}\;,
    \label{prg}
\ee
and $k_\perp(r_g,x) = \omega x(1-x) \tan{\frac{r_g}{2}}$. 
The power of this observable is that it is sensitive to the medium modification of the hardest branching inside jets, rather than the soft radiation which
can be transported to larger angles through different mechanisms such QGP excitations.
In FIG. \ref{Radius} we predict the angular separation modification for the leading subjets in the SCET$_{\rm G}$ framework. The same jet selection cuts and soft drop parameters are used as in the preliminary CMS momentum sharing measurements. We examine the $p_T$ dependence of the  angular region where the distribution is enhanced shifts to smaller values when the jet $p_T$ increases.  The  peak of this distribution corresponds to
the characteristic $r_g$  where the medium enhancement of large-angle splitting for hard branching processes is most significant.

To conclude, we presented the first calculation of the momentum sharing distribution $p(z_g)$, defined in the soft drop jet grooming procedure, and examined its modification in heavy ion collisions. This observable probes the hard branching at the early stage of parton shower formation and is a new powerful way to investigate the jet formation mechanism. In heavy ion collisions, the momentum sharing distribution of the two leading subjets in a reconstructed jets allows us to probe the early stages of the QGP evolution. We found that the $z_g$ distribution is significantly modified in the medium, as shown in our theory calculation and the preliminary CMS data. This suggests that the parton shower modification in the QGP starts early with the first hard splittings, and the in-medium splitting functions have a qualitatively different behavior from the ones in the vacuum. We also proposed a new measurement of the angular separation distribution between the leading subjets inside a groomed jet which encodes the angular distribution of the hard splitting, and we present theoretical predictions for its behavior. Future studies of jet substructure observables  more sensitive to the soft radiation, for example the jet mass
\cite{Chien:2010kc,Chien:2012ur,Chien:2016bnl,Kolodrubetz:2016dzb,Idilbi:2016hoa},
will allow us to map out the whole jet formation history.


Y.-T. C. would like to thank Andrew Larkoski, Yen-Jie Lee, Yacine Mehtar-Tani, Felix Ringer, Jesse Thaler and Marta Verweij for very helpful discussions, and the ESI at Universit{\"a}t Wien for hospitality and support. This research was supported by the US Department of Energy, Office of Science under Contract No. DE-AC52-06NA25396 and the DOE Early Career Program.

\bibliography{jet_shape}

\begin{thebibliography}{53}
\expandafter\ifx\csname natexlab\endcsname\relax\def\natexlab#1{#1}\fi
\expandafter\ifx\csname bibnamefont\endcsname\relax
  \def\bibnamefont#1{#1}\fi
\expandafter\ifx\csname bibfnamefont\endcsname\relax
  \def\bibfnamefont#1{#1}\fi
\expandafter\ifx\csname citenamefont\endcsname\relax
  \def\citenamefont#1{#1}\fi
\expandafter\ifx\csname url\endcsname\relax
  \def\url#1{\texttt{#1}}\fi
\expandafter\ifx\csname urlprefix\endcsname\relax\def\urlprefix{URL }\fi
\providecommand{\bibinfo}[2]{#2}
\providecommand{\eprint}[2][]{\url{#2}}

\bibitem[{\citenamefont{Adcox et~al.}(2002)}]{Adcox:2001jp}
\bibinfo{author}{\bibfnamefont{K.}~\bibnamefont{Adcox}} \bibnamefont{et~al.}
  (\bibinfo{collaboration}{PHENIX}), \bibinfo{journal}{Phys. Rev. Lett.}
  \textbf{\bibinfo{volume}{88}}, \bibinfo{pages}{022301}
  (\bibinfo{year}{2002}), \eprint{nucl-ex/0109003}.

\bibitem[{\citenamefont{Adler et~al.}(2002)}]{Adler:2002xw}
\bibinfo{author}{\bibfnamefont{C.}~\bibnamefont{Adler}} \bibnamefont{et~al.}
  (\bibinfo{collaboration}{STAR}), \bibinfo{journal}{Phys. Rev. Lett.}
  \textbf{\bibinfo{volume}{89}}, \bibinfo{pages}{202301}
  (\bibinfo{year}{2002}), \eprint{nucl-ex/0206011}.

\bibitem[{\citenamefont{Adcox et~al.}(2005)}]{Adcox:2004mh}
\bibinfo{author}{\bibfnamefont{K.}~\bibnamefont{Adcox}} \bibnamefont{et~al.}
  (\bibinfo{collaboration}{PHENIX}), \bibinfo{journal}{Nucl. Phys.}
  \textbf{\bibinfo{volume}{A757}}, \bibinfo{pages}{184} (\bibinfo{year}{2005}),
  \eprint{nucl-ex/0410003}.

\bibitem[{\citenamefont{Arsene et~al.}(2005)}]{Arsene:2004fa}
\bibinfo{author}{\bibfnamefont{I.}~\bibnamefont{Arsene}} \bibnamefont{et~al.}
  (\bibinfo{collaboration}{BRAHMS}), \bibinfo{journal}{Nucl. Phys.}
  \textbf{\bibinfo{volume}{A757}}, \bibinfo{pages}{1} (\bibinfo{year}{2005}),
  \eprint{nucl-ex/0410020}.

\bibitem[{\citenamefont{Back et~al.}(2005)}]{Back:2004je}
\bibinfo{author}{\bibfnamefont{B.~B.} \bibnamefont{Back}} \bibnamefont{et~al.},
  \bibinfo{journal}{Nucl. Phys.} \textbf{\bibinfo{volume}{A757}},
  \bibinfo{pages}{28} (\bibinfo{year}{2005}), \eprint{nucl-ex/0410022}.

\bibitem[{\citenamefont{Adams et~al.}(2005)}]{Adams:2005dq}
\bibinfo{author}{\bibfnamefont{J.}~\bibnamefont{Adams}} \bibnamefont{et~al.}
  (\bibinfo{collaboration}{STAR}), \bibinfo{journal}{Nucl. Phys.}
  \textbf{\bibinfo{volume}{A757}}, \bibinfo{pages}{102} (\bibinfo{year}{2005}),
  \eprint{nucl-ex/0501009}.

\bibitem[{\citenamefont{Aad et~al.}(2010)}]{Aad:2010bu}
\bibinfo{author}{\bibfnamefont{G.}~\bibnamefont{Aad}} \bibnamefont{et~al.}
  (\bibinfo{collaboration}{ATLAS}), \bibinfo{journal}{Phys.Rev.Lett.}
  \textbf{\bibinfo{volume}{105}}, \bibinfo{pages}{252303}
  (\bibinfo{year}{2010}), \eprint{1011.6182}.

\bibitem[{\citenamefont{Aamodt et~al.}(2011)}]{Aamodt:2010jd}
\bibinfo{author}{\bibfnamefont{K.}~\bibnamefont{Aamodt}} \bibnamefont{et~al.}
  (\bibinfo{collaboration}{ALICE}), \bibinfo{journal}{Phys. Lett.}
  \textbf{\bibinfo{volume}{B696}}, \bibinfo{pages}{30} (\bibinfo{year}{2011}),
  \eprint{1012.1004}.

\bibitem[{\citenamefont{Chatrchyan et~al.}(2011)}]{Chatrchyan:2011sx}
\bibinfo{author}{\bibfnamefont{S.}~\bibnamefont{Chatrchyan}}
  \bibnamefont{et~al.} (\bibinfo{collaboration}{CMS}),
  \bibinfo{journal}{Phys.Rev.} \textbf{\bibinfo{volume}{C84}},
  \bibinfo{pages}{024906} (\bibinfo{year}{2011}), \eprint{1102.1957}.

\bibitem[{\citenamefont{Chatrchyan et~al.}(2012{\natexlab{a}})}]{CMS:2012aa}
\bibinfo{author}{\bibfnamefont{S.}~\bibnamefont{Chatrchyan}}
  \bibnamefont{et~al.} (\bibinfo{collaboration}{CMS}), \bibinfo{journal}{Eur.
  Phys. J.} \textbf{\bibinfo{volume}{C72}}, \bibinfo{pages}{1945}
  (\bibinfo{year}{2012}{\natexlab{a}}), \eprint{1202.2554}.

\bibitem[{\citenamefont{Aad et~al.}(2013)}]{Aad:2012vca}
\bibinfo{author}{\bibfnamefont{G.}~\bibnamefont{Aad}} \bibnamefont{et~al.}
  (\bibinfo{collaboration}{ATLAS}), \bibinfo{journal}{Phys.Lett.}
  \textbf{\bibinfo{volume}{B719}}, \bibinfo{pages}{220} (\bibinfo{year}{2013}),
  \eprint{1208.1967}.

\bibitem[{\citenamefont{Aad et~al.}(2015)}]{Aad:2014bxa}
\bibinfo{author}{\bibfnamefont{G.}~\bibnamefont{Aad}} \bibnamefont{et~al.}
  (\bibinfo{collaboration}{ATLAS}), \bibinfo{journal}{Phys. Rev. Lett.}
  \textbf{\bibinfo{volume}{114}}, \bibinfo{pages}{072302}
  (\bibinfo{year}{2015}), \eprint{1411.2357}.

\bibitem[{\citenamefont{Chatrchyan et~al.}(2012{\natexlab{b}})}]{CMS:prelim}
\bibinfo{author}{\bibfnamefont{S.}~\bibnamefont{Chatrchyan}}
  \bibnamefont{et~al.} (\bibinfo{collaboration}{CMS})
  (\bibinfo{year}{2012}{\natexlab{b}}), \eprint{CMS-HIN-12-004}.

\bibitem[{\citenamefont{Adam et~al.}(2015)}]{Adam:2015ewa}
\bibinfo{author}{\bibfnamefont{J.}~\bibnamefont{Adam}} \bibnamefont{et~al.}
  (\bibinfo{collaboration}{ALICE}), \bibinfo{journal}{Phys. Lett.}
  \textbf{\bibinfo{volume}{B746}}, \bibinfo{pages}{1} (\bibinfo{year}{2015}),
  \eprint{1502.01689}.

\bibitem[{\citenamefont{Chatrchyan
  et~al.}(2014{\natexlab{a}})}]{Chatrchyan:2013kwa}
\bibinfo{author}{\bibfnamefont{S.}~\bibnamefont{Chatrchyan}}
  \bibnamefont{et~al.} (\bibinfo{collaboration}{CMS}),
  \bibinfo{journal}{Phys.Lett.} \textbf{\bibinfo{volume}{B730}},
  \bibinfo{pages}{243} (\bibinfo{year}{2014}{\natexlab{a}}),
  \eprint{1310.0878}.

\bibitem[{\citenamefont{Chatrchyan
  et~al.}(2014{\natexlab{b}})}]{Chatrchyan:2014ava}
\bibinfo{author}{\bibfnamefont{S.}~\bibnamefont{Chatrchyan}}
  \bibnamefont{et~al.} (\bibinfo{collaboration}{CMS}), \bibinfo{journal}{Phys.
  Rev.} \textbf{\bibinfo{volume}{C90}}, \bibinfo{pages}{024908}
  (\bibinfo{year}{2014}{\natexlab{b}}), \eprint{1406.0932}.

\bibitem[{\citenamefont{Aad et~al.}(2014)}]{Aad:2014wha}
\bibinfo{author}{\bibfnamefont{G.}~\bibnamefont{Aad}} \bibnamefont{et~al.}
  (\bibinfo{collaboration}{ATLAS}), \bibinfo{journal}{Phys. Lett.}
  \textbf{\bibinfo{volume}{B739}}, \bibinfo{pages}{320} (\bibinfo{year}{2014}),
  \eprint{1406.2979}.

\bibitem[{\citenamefont{Collaboration}(2016)}]{CMS:2016jys}
\bibinfo{author}{\bibfnamefont{C.}~\bibnamefont{Collaboration}}
  (\bibinfo{collaboration}{CMS}) (\bibinfo{year}{2016}).

\bibitem[{\citenamefont{Ellis et~al.}(1992)\citenamefont{Ellis, Kunszt, and
  Soper}}]{Ellis:1992qq}
\bibinfo{author}{\bibfnamefont{S.~D.} \bibnamefont{Ellis}},
  \bibinfo{author}{\bibfnamefont{Z.}~\bibnamefont{Kunszt}}, \bibnamefont{and}
  \bibinfo{author}{\bibfnamefont{D.~E.} \bibnamefont{Soper}},
  \bibinfo{journal}{Phys.Rev.Lett.} \textbf{\bibinfo{volume}{69}},
  \bibinfo{pages}{3615} (\bibinfo{year}{1992}), \eprint{hep-ph/9208249}.

\bibitem[{\citenamefont{Procura and Stewart}(2010)}]{Procura:2009vm}
\bibinfo{author}{\bibfnamefont{M.}~\bibnamefont{Procura}} \bibnamefont{and}
  \bibinfo{author}{\bibfnamefont{I.~W.} \bibnamefont{Stewart}},
  \bibinfo{journal}{Phys. Rev.} \textbf{\bibinfo{volume}{D81}},
  \bibinfo{pages}{074009} (\bibinfo{year}{2010}), \bibinfo{note}{[Erratum:
  Phys. Rev.D83,039902(2011)]}, \eprint{0911.4980}.

\bibitem[{\citenamefont{Chien and Vitev}(2014)}]{Chien:2014nsa}
\bibinfo{author}{\bibfnamefont{Y.-T.} \bibnamefont{Chien}} \bibnamefont{and}
  \bibinfo{author}{\bibfnamefont{I.}~\bibnamefont{Vitev}},
  \bibinfo{journal}{JHEP} \textbf{\bibinfo{volume}{12}}, \bibinfo{pages}{061}
  (\bibinfo{year}{2014}), \eprint{1405.4293}.

\bibitem[{\citenamefont{Chien and Vitev}(2016)}]{Chien:2015hda}
\bibinfo{author}{\bibfnamefont{Y.-T.} \bibnamefont{Chien}} \bibnamefont{and}
  \bibinfo{author}{\bibfnamefont{I.}~\bibnamefont{Vitev}},
  \bibinfo{journal}{JHEP} \textbf{\bibinfo{volume}{05}}, \bibinfo{pages}{023}
  (\bibinfo{year}{2016}), \eprint{1509.07257}.

\bibitem[{\citenamefont{Chien et~al.}(2016{\natexlab{a}})\citenamefont{Chien,
  Kang, Ringer, Vitev, and Xing}}]{Chien:2015ctp}
\bibinfo{author}{\bibfnamefont{Y.-T.} \bibnamefont{Chien}},
  \bibinfo{author}{\bibfnamefont{Z.-B.} \bibnamefont{Kang}},
  \bibinfo{author}{\bibfnamefont{F.}~\bibnamefont{Ringer}},
  \bibinfo{author}{\bibfnamefont{I.}~\bibnamefont{Vitev}}, \bibnamefont{and}
  \bibinfo{author}{\bibfnamefont{H.}~\bibnamefont{Xing}},
  \bibinfo{journal}{JHEP} \textbf{\bibinfo{volume}{05}}, \bibinfo{pages}{125}
  (\bibinfo{year}{2016}{\natexlab{a}}), \eprint{1512.06851}.

\bibitem[{\citenamefont{Chien et~al.}(2016{\natexlab{b}})\citenamefont{Chien,
  Kang, Ringer, Vitev, and Xing}}]{medium_jet_frag}
\bibinfo{author}{\bibfnamefont{Y.-T.} \bibnamefont{Chien}},
  \bibinfo{author}{\bibfnamefont{Z.-B.} \bibnamefont{Kang}},
  \bibinfo{author}{\bibfnamefont{F.}~\bibnamefont{Ringer}},
  \bibinfo{author}{\bibfnamefont{I.}~\bibnamefont{Vitev}}, \bibnamefont{and}
  \bibinfo{author}{\bibfnamefont{H.}~\bibnamefont{Xing}}
  (\bibinfo{year}{2016}{\natexlab{b}}), \eprint{in preparation}.

\bibitem[{\citenamefont{Larkoski et~al.}(2014)\citenamefont{Larkoski, Marzani,
  Soyez, and Thaler}}]{Larkoski:2014wba}
\bibinfo{author}{\bibfnamefont{A.~J.} \bibnamefont{Larkoski}},
  \bibinfo{author}{\bibfnamefont{S.}~\bibnamefont{Marzani}},
  \bibinfo{author}{\bibfnamefont{G.}~\bibnamefont{Soyez}}, \bibnamefont{and}
  \bibinfo{author}{\bibfnamefont{J.}~\bibnamefont{Thaler}},
  \bibinfo{journal}{JHEP} \textbf{\bibinfo{volume}{05}}, \bibinfo{pages}{146}
  (\bibinfo{year}{2014}), \eprint{1402.2657}.

\bibitem[{\citenamefont{Larkoski et~al.}(2015)\citenamefont{Larkoski, Marzani,
  and Thaler}}]{Larkoski:2015lea}
\bibinfo{author}{\bibfnamefont{A.~J.} \bibnamefont{Larkoski}},
  \bibinfo{author}{\bibfnamefont{S.}~\bibnamefont{Marzani}}, \bibnamefont{and}
  \bibinfo{author}{\bibfnamefont{J.}~\bibnamefont{Thaler}},
  \bibinfo{journal}{Phys. Rev.} \textbf{\bibinfo{volume}{D91}},
  \bibinfo{pages}{111501} (\bibinfo{year}{2015}), \eprint{1502.01719}.

\bibitem[{\citenamefont{Altarelli and Parisi}(1977)}]{Altarelli:1977zs}
\bibinfo{author}{\bibfnamefont{G.}~\bibnamefont{Altarelli}} \bibnamefont{and}
  \bibinfo{author}{\bibfnamefont{G.}~\bibnamefont{Parisi}},
  \bibinfo{journal}{Nucl. Phys.} \textbf{\bibinfo{volume}{B126}},
  \bibinfo{pages}{298} (\bibinfo{year}{1977}).

\bibitem[{\citenamefont{Cacciari et~al.}(2008)\citenamefont{Cacciari, Salam,
  and Soyez}}]{Cacciari:2008gp}
\bibinfo{author}{\bibfnamefont{M.}~\bibnamefont{Cacciari}},
  \bibinfo{author}{\bibfnamefont{G.~P.} \bibnamefont{Salam}}, \bibnamefont{and}
  \bibinfo{author}{\bibfnamefont{G.}~\bibnamefont{Soyez}},
  \bibinfo{journal}{JHEP} \textbf{\bibinfo{volume}{0804}}, \bibinfo{pages}{063}
  (\bibinfo{year}{2008}), \eprint{0802.1189}.

\bibitem[{\citenamefont{Dokshitzer et~al.}(1997)\citenamefont{Dokshitzer,
  Leder, Moretti, and Webber}}]{Dokshitzer:1997in}
\bibinfo{author}{\bibfnamefont{Y.~L.} \bibnamefont{Dokshitzer}},
  \bibinfo{author}{\bibfnamefont{G.~D.} \bibnamefont{Leder}},
  \bibinfo{author}{\bibfnamefont{S.}~\bibnamefont{Moretti}}, \bibnamefont{and}
  \bibinfo{author}{\bibfnamefont{B.~R.} \bibnamefont{Webber}},
  \bibinfo{journal}{JHEP} \textbf{\bibinfo{volume}{08}}, \bibinfo{pages}{001}
  (\bibinfo{year}{1997}), \eprint{hep-ph/9707323}.

\bibitem[{\citenamefont{Wobisch and Wengler}(1998)}]{Wobisch:1998wt}
\bibinfo{author}{\bibfnamefont{M.}~\bibnamefont{Wobisch}} \bibnamefont{and}
  \bibinfo{author}{\bibfnamefont{T.}~\bibnamefont{Wengler}}
  (\bibinfo{year}{1998}), \eprint{hep-ph/9907280},
  \urlprefix\url{https://inspirehep.net/record/484872/files/arXiv:hep-ph_9907280.pdf}.

\bibitem[{\citenamefont{Dasgupta et~al.}(2015)\citenamefont{Dasgupta, Dreyer,
  Salam, and Soyez}}]{Dasgupta:2014yra}
\bibinfo{author}{\bibfnamefont{M.}~\bibnamefont{Dasgupta}},
  \bibinfo{author}{\bibfnamefont{F.}~\bibnamefont{Dreyer}},
  \bibinfo{author}{\bibfnamefont{G.~P.} \bibnamefont{Salam}}, \bibnamefont{and}
  \bibinfo{author}{\bibfnamefont{G.}~\bibnamefont{Soyez}},
  \bibinfo{journal}{JHEP} \textbf{\bibinfo{volume}{04}}, \bibinfo{pages}{039}
  (\bibinfo{year}{2015}), \eprint{1411.5182}.

\bibitem[{\citenamefont{Chien et~al.}(2016{\natexlab{c}})\citenamefont{Chien,
  Hornig, and Lee}}]{Chien:2015cka}
\bibinfo{author}{\bibfnamefont{Y.-T.} \bibnamefont{Chien}},
  \bibinfo{author}{\bibfnamefont{A.}~\bibnamefont{Hornig}}, \bibnamefont{and}
  \bibinfo{author}{\bibfnamefont{C.}~\bibnamefont{Lee}},
  \bibinfo{journal}{Phys. Rev.} \textbf{\bibinfo{volume}{D93}},
  \bibinfo{pages}{014033} (\bibinfo{year}{2016}{\natexlab{c}}),
  \eprint{1509.04287}.

\bibitem[{\citenamefont{Becher et~al.}(2016)\citenamefont{Becher, Neubert,
  Rothen, and Shao}}]{Becher:2015hka}
\bibinfo{author}{\bibfnamefont{T.}~\bibnamefont{Becher}},
  \bibinfo{author}{\bibfnamefont{M.}~\bibnamefont{Neubert}},
  \bibinfo{author}{\bibfnamefont{L.}~\bibnamefont{Rothen}}, \bibnamefont{and}
  \bibinfo{author}{\bibfnamefont{D.~Y.} \bibnamefont{Shao}},
  \bibinfo{journal}{Phys. Rev. Lett.} \textbf{\bibinfo{volume}{116}},
  \bibinfo{pages}{192001} (\bibinfo{year}{2016}), \eprint{1508.06645}.

\bibitem[{\citenamefont{Kang et~al.}(2016)\citenamefont{Kang, Ringer, and
  Vitev}}]{Kang:2016mcy}
\bibinfo{author}{\bibfnamefont{Z.-B.} \bibnamefont{Kang}},
  \bibinfo{author}{\bibfnamefont{F.}~\bibnamefont{Ringer}}, \bibnamefont{and}
  \bibinfo{author}{\bibfnamefont{I.}~\bibnamefont{Vitev}}
  (\bibinfo{year}{2016}), \eprint{1606.06732}.

\bibitem[{\citenamefont{Bauer et~al.}(2000)\citenamefont{Bauer, Fleming, and
  Luke}}]{Bauer:2000ew}
\bibinfo{author}{\bibfnamefont{C.~W.} \bibnamefont{Bauer}},
  \bibinfo{author}{\bibfnamefont{S.}~\bibnamefont{Fleming}}, \bibnamefont{and}
  \bibinfo{author}{\bibfnamefont{M.~E.} \bibnamefont{Luke}},
  \bibinfo{journal}{Phys.Rev.} \textbf{\bibinfo{volume}{D63}},
  \bibinfo{pages}{014006} (\bibinfo{year}{2000}), \eprint{hep-ph/0005275}.

\bibitem[{\citenamefont{Bauer et~al.}(2001)\citenamefont{Bauer, Fleming,
  Pirjol, and Stewart}}]{Bauer:2000yr}
\bibinfo{author}{\bibfnamefont{C.~W.} \bibnamefont{Bauer}},
  \bibinfo{author}{\bibfnamefont{S.}~\bibnamefont{Fleming}},
  \bibinfo{author}{\bibfnamefont{D.}~\bibnamefont{Pirjol}}, \bibnamefont{and}
  \bibinfo{author}{\bibfnamefont{I.~W.} \bibnamefont{Stewart}},
  \bibinfo{journal}{Phys.Rev.} \textbf{\bibinfo{volume}{D63}},
  \bibinfo{pages}{114020} (\bibinfo{year}{2001}), \eprint{hep-ph/0011336}.

\bibitem[{\citenamefont{Bauer and Stewart}(2001)}]{Bauer:2001ct}
\bibinfo{author}{\bibfnamefont{C.~W.} \bibnamefont{Bauer}} \bibnamefont{and}
  \bibinfo{author}{\bibfnamefont{I.~W.} \bibnamefont{Stewart}},
  \bibinfo{journal}{Phys.Lett.} \textbf{\bibinfo{volume}{B516}},
  \bibinfo{pages}{134} (\bibinfo{year}{2001}), \eprint{hep-ph/0107001}.

\bibitem[{\citenamefont{Bauer et~al.}(2002{\natexlab{a}})\citenamefont{Bauer,
  Pirjol, and Stewart}}]{Bauer:2001yt}
\bibinfo{author}{\bibfnamefont{C.~W.} \bibnamefont{Bauer}},
  \bibinfo{author}{\bibfnamefont{D.}~\bibnamefont{Pirjol}}, \bibnamefont{and}
  \bibinfo{author}{\bibfnamefont{I.~W.} \bibnamefont{Stewart}},
  \bibinfo{journal}{Phys.Rev.} \textbf{\bibinfo{volume}{D65}},
  \bibinfo{pages}{054022} (\bibinfo{year}{2002}{\natexlab{a}}),
  \eprint{hep-ph/0109045}.

\bibitem[{\citenamefont{Bauer et~al.}(2002{\natexlab{b}})\citenamefont{Bauer,
  Fleming, Pirjol, Rothstein, and Stewart}}]{Bauer:2002nz}
\bibinfo{author}{\bibfnamefont{C.~W.} \bibnamefont{Bauer}},
  \bibinfo{author}{\bibfnamefont{S.}~\bibnamefont{Fleming}},
  \bibinfo{author}{\bibfnamefont{D.}~\bibnamefont{Pirjol}},
  \bibinfo{author}{\bibfnamefont{I.~Z.} \bibnamefont{Rothstein}},
  \bibnamefont{and} \bibinfo{author}{\bibfnamefont{I.~W.}
  \bibnamefont{Stewart}}, \bibinfo{journal}{Phys.Rev.}
  \textbf{\bibinfo{volume}{D66}}, \bibinfo{pages}{014017}
  (\bibinfo{year}{2002}{\natexlab{b}}), \eprint{hep-ph/0202088}.

\bibitem[{\citenamefont{Idilbi and Majumder}(2009)}]{Idilbi:2008vm}
\bibinfo{author}{\bibfnamefont{A.}~\bibnamefont{Idilbi}} \bibnamefont{and}
  \bibinfo{author}{\bibfnamefont{A.}~\bibnamefont{Majumder}},
  \bibinfo{journal}{Phys.Rev.} \textbf{\bibinfo{volume}{D80}},
  \bibinfo{pages}{054022} (\bibinfo{year}{2009}), \eprint{0808.1087}.

\bibitem[{\citenamefont{Ovanesyan and Vitev}(2011)}]{Ovanesyan:2011xy}
\bibinfo{author}{\bibfnamefont{G.}~\bibnamefont{Ovanesyan}} \bibnamefont{and}
  \bibinfo{author}{\bibfnamefont{I.}~\bibnamefont{Vitev}},
  \bibinfo{journal}{JHEP} \textbf{\bibinfo{volume}{1106}}, \bibinfo{pages}{080}
  (\bibinfo{year}{2011}), \eprint{1103.1074}.

\bibitem[{\citenamefont{Ovanesyan and Vitev}(2012)}]{Ovanesyan:2011kn}
\bibinfo{author}{\bibfnamefont{G.}~\bibnamefont{Ovanesyan}} \bibnamefont{and}
  \bibinfo{author}{\bibfnamefont{I.}~\bibnamefont{Vitev}},
  \bibinfo{journal}{Phys.Lett.} \textbf{\bibinfo{volume}{B706}},
  \bibinfo{pages}{371} (\bibinfo{year}{2012}), \eprint{1109.5619}.

\bibitem[{\citenamefont{Fickinger et~al.}(2013)\citenamefont{Fickinger,
  Ovanesyan, and Vitev}}]{Fickinger:2013xwa}
\bibinfo{author}{\bibfnamefont{M.}~\bibnamefont{Fickinger}},
  \bibinfo{author}{\bibfnamefont{G.}~\bibnamefont{Ovanesyan}},
  \bibnamefont{and} \bibinfo{author}{\bibfnamefont{I.}~\bibnamefont{Vitev}},
  \bibinfo{journal}{JHEP} \textbf{\bibinfo{volume}{1307}}, \bibinfo{pages}{059}
  (\bibinfo{year}{2013}), \eprint{1304.3497}.

\bibitem[{\citenamefont{Bjorken}(1982)}]{Bjorken:1982tu}
\bibinfo{author}{\bibfnamefont{J.~D.} \bibnamefont{Bjorken}}, pp.
  \bibinfo{pages}{FERMILAB--PUB--82--059--THY, FERMILAB--PUB--82--059--T}
  (\bibinfo{year}{1982}).

\bibitem[{\citenamefont{Kang et~al.}(2014)\citenamefont{Kang, Lashof-Regas,
  Ovanesyan, Saad, and Vitev}}]{Kang:2014xsa}
\bibinfo{author}{\bibfnamefont{Z.-B.} \bibnamefont{Kang}},
  \bibinfo{author}{\bibfnamefont{R.}~\bibnamefont{Lashof-Regas}},
  \bibinfo{author}{\bibfnamefont{G.}~\bibnamefont{Ovanesyan}},
  \bibinfo{author}{\bibfnamefont{P.}~\bibnamefont{Saad}}, \bibnamefont{and}
  \bibinfo{author}{\bibfnamefont{I.}~\bibnamefont{Vitev}}
  (\bibinfo{year}{2014}), \eprint{1405.2612}.

\bibitem[{\citenamefont{Chien et~al.}(2016{\natexlab{d}})\citenamefont{Chien,
  Emerman, Kang, Ovanesyan, and Vitev}}]{Chien:2015vja}
\bibinfo{author}{\bibfnamefont{Y.-T.} \bibnamefont{Chien}},
  \bibinfo{author}{\bibfnamefont{A.}~\bibnamefont{Emerman}},
  \bibinfo{author}{\bibfnamefont{Z.-B.} \bibnamefont{Kang}},
  \bibinfo{author}{\bibfnamefont{G.}~\bibnamefont{Ovanesyan}},
  \bibnamefont{and} \bibinfo{author}{\bibfnamefont{I.}~\bibnamefont{Vitev}},
  \bibinfo{journal}{Phys. Rev.} \textbf{\bibinfo{volume}{D93}},
  \bibinfo{pages}{074030} (\bibinfo{year}{2016}{\natexlab{d}}),
  \eprint{1509.02936}.

\bibitem[{\citenamefont{Vitev}(2007)}]{Vitev:2007ve}
\bibinfo{author}{\bibfnamefont{I.}~\bibnamefont{Vitev}},
  \bibinfo{journal}{Phys. Rev.} \textbf{\bibinfo{volume}{C75}},
  \bibinfo{pages}{064906} (\bibinfo{year}{2007}), \eprint{hep-ph/0703002}.

\bibitem[{\citenamefont{Tung et~al.}(2007)\citenamefont{Tung, Lai, Belyaev,
  Pumplin, Stump et~al.}}]{Tung:2006tb}
\bibinfo{author}{\bibfnamefont{W.}~\bibnamefont{Tung}},
  \bibinfo{author}{\bibfnamefont{H.}~\bibnamefont{Lai}},
  \bibinfo{author}{\bibfnamefont{A.}~\bibnamefont{Belyaev}},
  \bibinfo{author}{\bibfnamefont{J.}~\bibnamefont{Pumplin}},
  \bibinfo{author}{\bibfnamefont{D.}~\bibnamefont{Stump}},
  \bibnamefont{et~al.}, \bibinfo{journal}{JHEP}
  \textbf{\bibinfo{volume}{0702}}, \bibinfo{pages}{053} (\bibinfo{year}{2007}),
  \eprint{hep-ph/0611254}.

\bibitem[{\citenamefont{Chien and Schwartz}(2010)}]{Chien:2010kc}
\bibinfo{author}{\bibfnamefont{Y.-T.} \bibnamefont{Chien}} \bibnamefont{and}
  \bibinfo{author}{\bibfnamefont{M.~D.} \bibnamefont{Schwartz}},
  \bibinfo{journal}{JHEP} \textbf{\bibinfo{volume}{1008}}, \bibinfo{pages}{058}
  (\bibinfo{year}{2010}), \eprint{1005.1644}.

\bibitem[{\citenamefont{Chien et~al.}(2013)\citenamefont{Chien, Kelley,
  Schwartz, and Zhu}}]{Chien:2012ur}
\bibinfo{author}{\bibfnamefont{Y.-T.} \bibnamefont{Chien}},
  \bibinfo{author}{\bibfnamefont{R.}~\bibnamefont{Kelley}},
  \bibinfo{author}{\bibfnamefont{M.~D.} \bibnamefont{Schwartz}},
  \bibnamefont{and} \bibinfo{author}{\bibfnamefont{H.~X.} \bibnamefont{Zhu}},
  \bibinfo{journal}{Phys. Rev.} \textbf{\bibinfo{volume}{D87}},
  \bibinfo{pages}{014010} (\bibinfo{year}{2013}), \eprint{1208.0010}.

\bibitem[{\citenamefont{Chien}(2016)}]{Chien:2016bnl}
\bibinfo{author}{\bibfnamefont{Y.-T.} \bibnamefont{Chien}},
  \bibinfo{journal}{Talk presented at the 11th International Workshop on
  High-$p_T$ Physics in the RHIC \& LHC Era and paper in preparation}
  \textbf{\bibinfo{volume}{https://indico.bnl.gov/get\\File.py/access?contribId=39\&sessionId=11\&resId\\=0\&materialId=slides\&confId=1800}}
  (\bibinfo{year}{2016}).

\bibitem[{\citenamefont{Kolodrubetz et~al.}(2016)\citenamefont{Kolodrubetz,
  Pietrulewicz, Stewart, Tackmann, and Waalewijn}}]{Kolodrubetz:2016dzb}
\bibinfo{author}{\bibfnamefont{D.~W.} \bibnamefont{Kolodrubetz}},
  \bibinfo{author}{\bibfnamefont{P.}~\bibnamefont{Pietrulewicz}},
  \bibinfo{author}{\bibfnamefont{I.~W.} \bibnamefont{Stewart}},
  \bibinfo{author}{\bibfnamefont{F.~J.} \bibnamefont{Tackmann}},
  \bibnamefont{and} \bibinfo{author}{\bibfnamefont{W.~J.}
  \bibnamefont{Waalewijn}} (\bibinfo{year}{2016}), \eprint{1605.08038}.

\bibitem[{\citenamefont{Idilbi and Kim}(2016)}]{Idilbi:2016hoa}
\bibinfo{author}{\bibfnamefont{A.}~\bibnamefont{Idilbi}} \bibnamefont{and}
  \bibinfo{author}{\bibfnamefont{C.}~\bibnamefont{Kim}} (\bibinfo{year}{2016}),
  \eprint{1606.05429}.

\end{thebibliography}
\end{document}